\documentclass[iop,appendixfloats,numberedappendix]{emulateapj}

\shorttitle{Abundance Patterns of $z\sim1.4$ Galaxies}
\shortauthors{Kriek et al.}
\usepackage{natbib}
\def\bs{\hspace{-0.06in}}
\slugcomment{Accepted for Publication in ApJL}

\begin{document}
  
\title{Stellar Metallicities and Elemental Abundance Ratios of $\lowercase{z}\sim 1.4$ Massive Quiescent Galaxies\altaffilmark{*}}

\author{Mariska Kriek\altaffilmark{1}, Sedona H. Price\altaffilmark{2}, Charlie Conroy\altaffilmark{3}, Katherine Suess\altaffilmark{1}, Lamiya Mowla\altaffilmark{4}, Imad Pasha\altaffilmark{4}, Rachel Bezanson\altaffilmark{5}, Pieter van Dokkum\altaffilmark{4} \& Guillermo Barro\altaffilmark{6}}

\altaffiltext{*}{Based on data obtained at the W.M. Keck Observatory
  and with the NASA/ESA Hubble Space Telescope (HST). W.M. Keck
  Observatory is operated as a scientific partnership among the
  California Institute of Technology, the University of California and
  NASA, and was made possible by the generous financial support of the
  W.M. Keck Foundation. HST is operated by the Association of
  Universities for Research in Astronomy, Inc., under NASA contract
  NAS 5-26555.}

\altaffiltext{1}{Astronomy Department, University of California, Berkeley, CA 94720, USA}

\altaffiltext{2}{Max Planck Institute for Extraterrestrial Physics, Giessenbachstrasse 1, D-85741 Garching, Germany}

\altaffiltext{3}{Harvard-Smithsonian Center for Astrophysics, 60 Garden St., Cambridge, MA, USA}

\altaffiltext{4}{Department of Astronomy, Yale University, New Haven, CT 06511, USA}

\altaffiltext{5}{Department of Physics and Astronomy and PITT PACC, University of Pittsburgh, Pittsburgh, PA, 15260, USA}

\altaffiltext{6}{Department of Physics, University of the Pacific, 3601 Pacific Avenue, Stockton, CA 95211, USA}

\begin{abstract}
The chemical composition of galaxies has been measured out to
$z\sim4$. However, nearly all studies beyond $z\sim0.7$ are based on
strong-line emission from HII regions within star-forming
galaxies. Measuring the chemical composition of distant quiescent
galaxies is extremely challenging, as the required stellar absorption
features are faint and shifted to near-infrared wavelengths. Here, we
present ultra-deep rest-frame optical spectra of five massive
quiescent galaxies at $z\sim1.4$, all of which show numerous stellar
absorption lines. We derive the abundance ratios [Mg/Fe] and [Fe/H]
for three out of five galaxies; the remaining two galaxies have too
young luminosity-weighted ages to yield robust measurements. Similar
to lower-redshift findings, [Mg/Fe] appears positively correlated with
stellar mass, while [Fe/H] is approximately constant with mass. These
results may imply that the stellar mass-metallicity relation was
already in place at $z\sim1.4$. While the [Mg/Fe]-mass relation at
$z\sim1.4$ is consistent with the $z<0.7$ relation, [Fe/H] at
$z\sim1.4$ is $\sim0.2$~dex lower than at $z<0.7$. With a [Mg/Fe] of
$0.44^{+0.08}_{-0.07}$ the most massive galaxy may be more
$\alpha$-enhanced than similar-mass galaxies at lower redshift, but
the offset is less significant than the [Mg/Fe] of 0.6 previously
found for a massive galaxy at $z=2.1$. Nonetheless, these results
combined may suggest that [Mg/Fe] in the most massive galaxies
decreases over time, possibly by accreting low-mass, less
$\alpha$-enhanced galaxies. A larger galaxy sample is needed to
confirm this scenario. Finally, the abundance ratios indicate short
star-formation timescales of $0.2-1.0$~Gyr.
  
\end{abstract} 


\section{INTRODUCTION}

The chemical composition of a galaxy reflects the interplay of several
fundamental physical processes in galaxy formation, including star
formation, metal production, feedback and gas exchange with the
surrounding medium, and galaxy merging. Subsequent star-formation
episodes and the recycling of enriched gas result in an increase of
metallicity with time, while the duration of the star-forming phase
sets the relative abundances of different metals. Feedback processes
further impact the chemical enrichment history, as they may expel
enriched gas from galaxies. In combination with inflow of lower
metallicity gas from the inter/circumgalactic medium, they alter the metal
content of the gas supply. Finally, galaxy mergers affect the chemical
composition, as stars in accreted galaxies have their own chemical
footprint \citep[see][and references therein]{RMaiolino2019}.

Metallicities of low-redshift galaxies have been studied extensively
for both the interstellar gas \citep[e.g.,][]{CTremonti2004} as well
as the gas locked up in stars \citep[e.g.,][]{AGallazzi2005}. To
disentangle the effects of star formation, feedback/gas exchange, and
galaxy mergers on the chemical abundance patterns of galaxies, it is
crucial to extend these studies to higher redshifts. Metallicities of
star-forming galaxies have been studied out to $z\sim4$
\citep[e.g.,][]{DErb2006a,FMannucci2009,AShapley2017}. However, the
chemical composition of quiescent galaxies has only been routinely
measured out to $z\sim0.7$
\citep[e.g.,][]{AGallazzi2014,JChoi2014}. Because quiescent galaxies
dominate the massive galaxy population out to $z\sim2$
\citep{AMuzzin2013b,ATomczak2014}, our current understanding of the
chemical composition of galaxies through cosmic time is thus
incomplete.

\begin{figure*}
\centering
\includegraphics[width=0.4\textwidth]{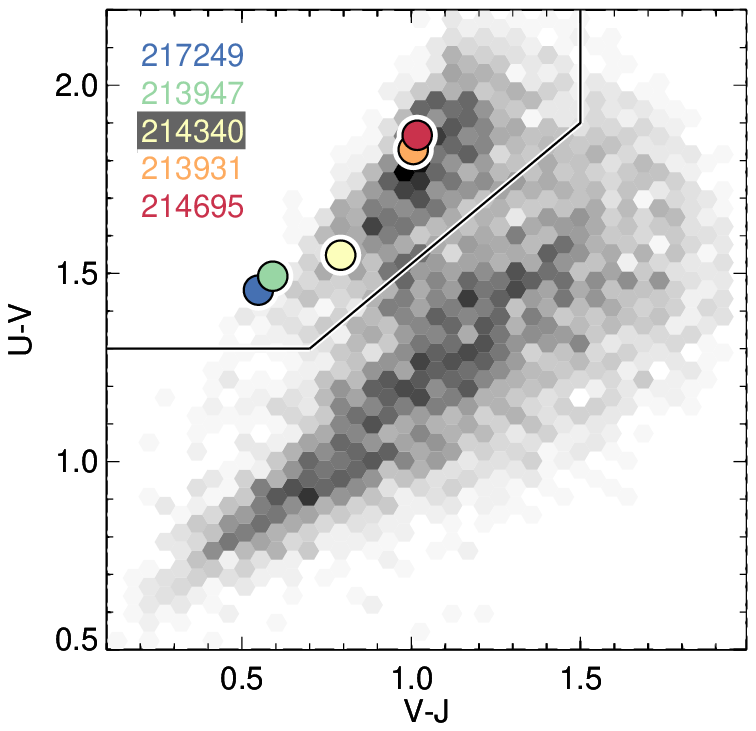}\hspace{0.2in}
\includegraphics[width=0.4\textwidth]{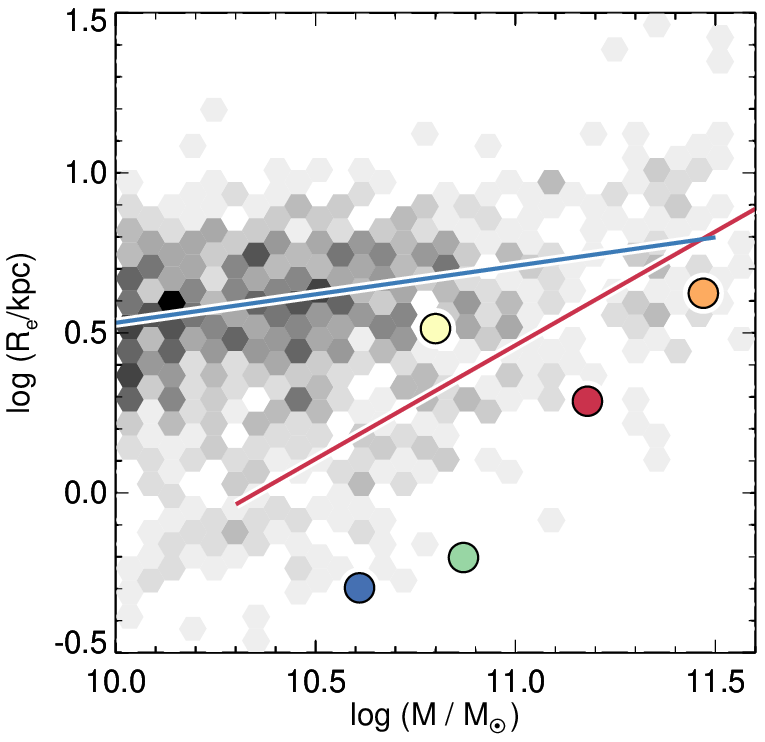}

\caption{{\it Left:} Rest-frame $U-V$ vs. $V-J$ of the five quiescent
  galaxies observed with MOSFIRE and LRIS in comparison to all
  galaxies with $1.3<z<1.5$ and log~($M/M_\odot)>10$ in the UltraVISTA
  field. {\it Right:} Rest-frame optical size (at 5000\,\AA) vs
  stellar mass of the same five galaxies in comparison to all galaxies
  with $1.3<z<1.5$ and log $(M/M_\odot)>10$ from
  \cite{LMowla2018}. The blue and red lines represent the best-fit
  relations for star-forming and quiescent galaxies, respectively, at
  $z\sim1.4$ \citep{LMowla2018}. The most massive galaxy, 213931,
    consists of three clumps. In this panel we show the size and
    mass of the brightest clump, where the mass has been estimated
    using the magnitude ratios of the clumps.}

\label{fig:uvj}
\end{figure*}

Measuring chemical compositions of $z>0.7$ quiescent galaxies is
extremely challenging as the required stellar absorption lines are
faint and shifted to near-infrared wavelengths. Furthermore, because
of the younger stellar ages, the metal absorption lines for distant
quiescent galaxies are weaker than for their local analogs. The few
available measurements at $z>0.7$ are either based on a stacked
spectrum of quiescent galaxies \citep[e.g.,][]{MOnodera2015}, or on an
ultradeep spectrum of a single quiescent galaxy
\citep[e.g.,][]{SToft2012,ILonoce2015,MKriek2016}. There are also several
studies that use low-resolution grism spectra obtained with the {\it
  Hubble Space Telescope (HST)} to derive stellar metallicities
\citep[e.g.,][]{TMorishita2018,VEstrada2018}. However, because these
studies primarily rely on the continuum shape, they are especially
susceptible to modeling degeneracies and other systematic errors.   

Early results suggest that $z>0.7$ quiescent galaxies have super-solar
metallicities \citep{MOnodera2015,ILonoce2015,MKriek2016} and are
$\alpha$-enhanced, with [Mg/Fe] that are similar \citep{MOnodera2015}
or significantly higher \citep{ILonoce2015,MKriek2016} than those of
similar-mass low-redshift galaxies. These high [Mg/Fe] imply very
short star-formation timescales, though the stellar initial mass
function (IMF) may also affect this abundance ratio
\citep[e.g.,][]{FFontanot2017}.  Furthermore, these measurements raise
the question of how high-redshift quiescent galaxies with high [Mg/Fe]
evolve into the early-type galaxy population with lower [Mg/Fe] seen
today. Larger galaxy samples are needed to confirm these results

In this Letter we present elemental abundance ratios for five massive
galaxies at $z\sim1.4$, derived from deep spectra obtained
with LRIS and MOSFIRE on the Keck I Telescope. This study is enabled
by the large UltraVISTA \citep{HMcCracken2012} and COSMOS-DASH
\citep{IMomcheva2017,LMowla2018} field, which facilitated the
identification of pointings for which we can observe several
bright targets (with HST/F160W imaging) simultaneously. Throughout this
work we assume a $\Lambda$CDM cosmology with $\Omega_{\textrm{m}}=0.3$,
$\Omega_{\rm\Lambda}=0.7$, and $H_0=70$~km\,s$^{-1}$\,Mpc$^{-1}$.

\section{GALAXY SAMPLE AND DATA}

The observed galaxies were identified using the UltraVISTA $K$-band
selected catalog (v4.1) by \cite{AMuzzin2013a}. We selected the
pointing for which we could observe the most $J<21.6$ quiescent
galaxies at $1.3<z<1.5$ in one MOSFIRE/LRIS mask. For this redshift
range, we catch prominent metal and Balmer absorption lines in
atmospheric windows. Galaxies were classified as quiescent based on
their rest-frame $U-V$ and $V-J$ colors
\citep[e.g.,][]{SWuyts2007}. Furthermore, we required that the
pointing overlaps with the COSMOS-DASH survey
\citep{IMomcheva2017,LMowla2018}, which provides shallow F160W imaging
and thus allows the measurement of rest-frame optical
sizes. Figure~\ref{fig:uvj} shows the location of the five targets in
the rest-frame $U-V$ vs. $V-J$ (UVJ) diagram, as well as in  $R_e$
vs. stellar mass space, compared to the full galaxy distribution with
$M>10^{10}\,M_\odot$ and $1.3<z<1.5$
\citep{AMuzzin2013b,LMowla2018}. The galaxies span a range in colors
along the quiescent sequence in the UVJ diagram, as well as a range in
sizes. On average, they are slightly bluer and smaller than the
typical quiescent galaxy at this redshift. This bias may be
  expected; our magnitude selection favors post-starburst galaxies,
  which are brighter, bluer, and presumably smaller than older
  quiescent galaxies of similar mass
  \citep[e.g.,][]{KWhitaker2012,MYano2016,OAlmaini2017}.

\begin{figure*}
  \centering
  \includegraphics[width=0.95\textwidth]{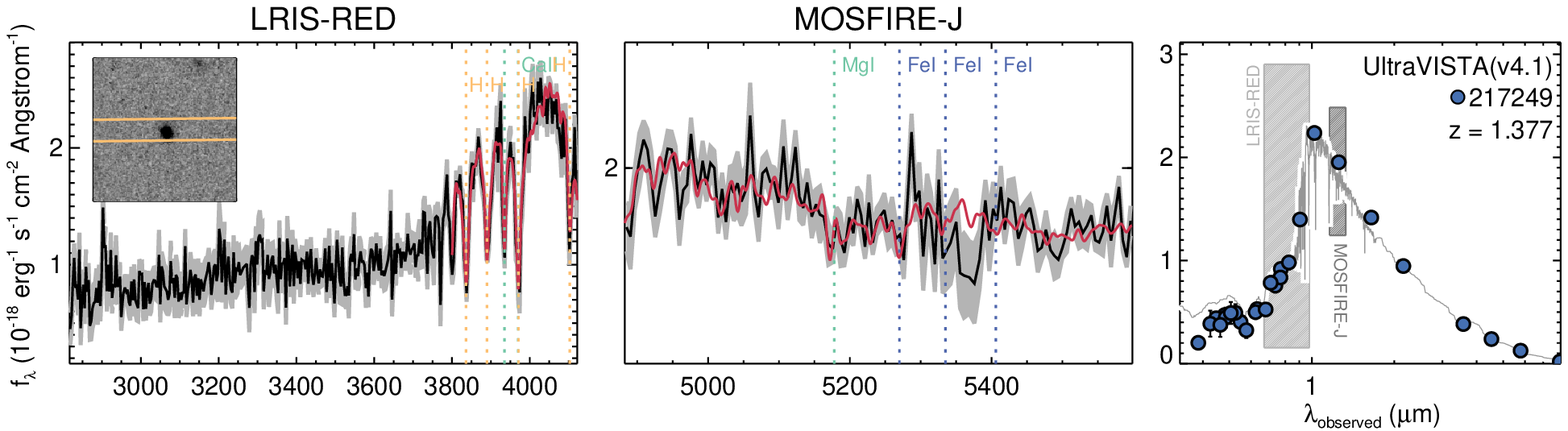}
  \includegraphics[width=0.95\textwidth]{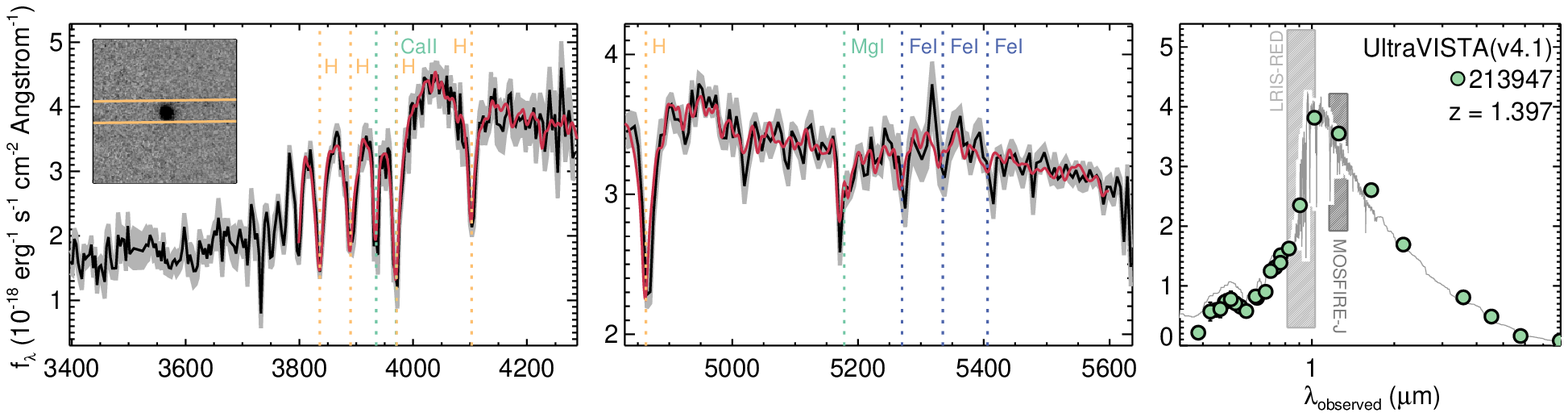}
  \includegraphics[width=0.95\textwidth]{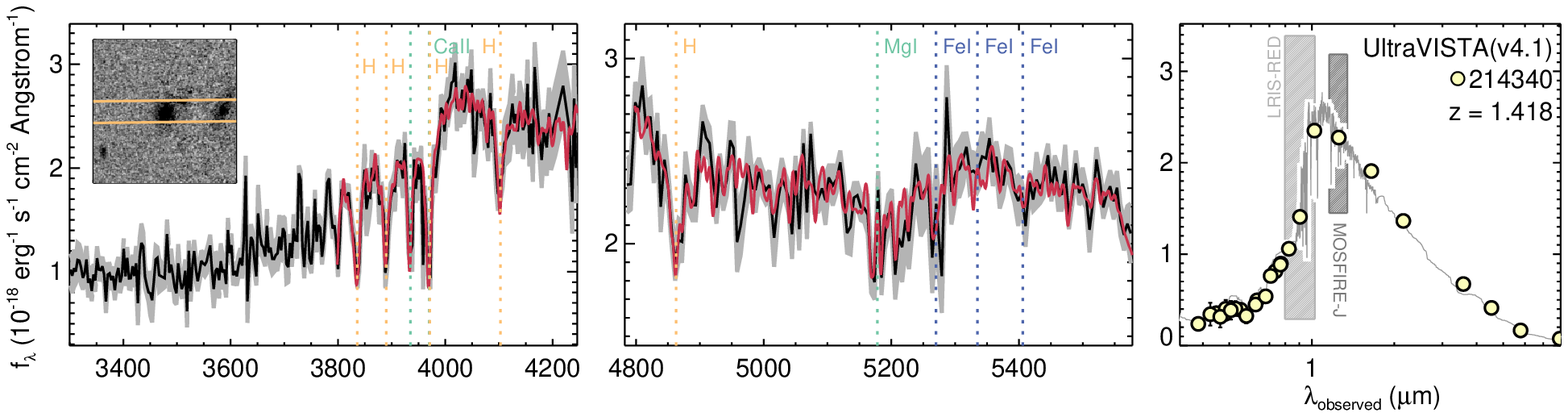}  
  \includegraphics[width=0.95\textwidth]{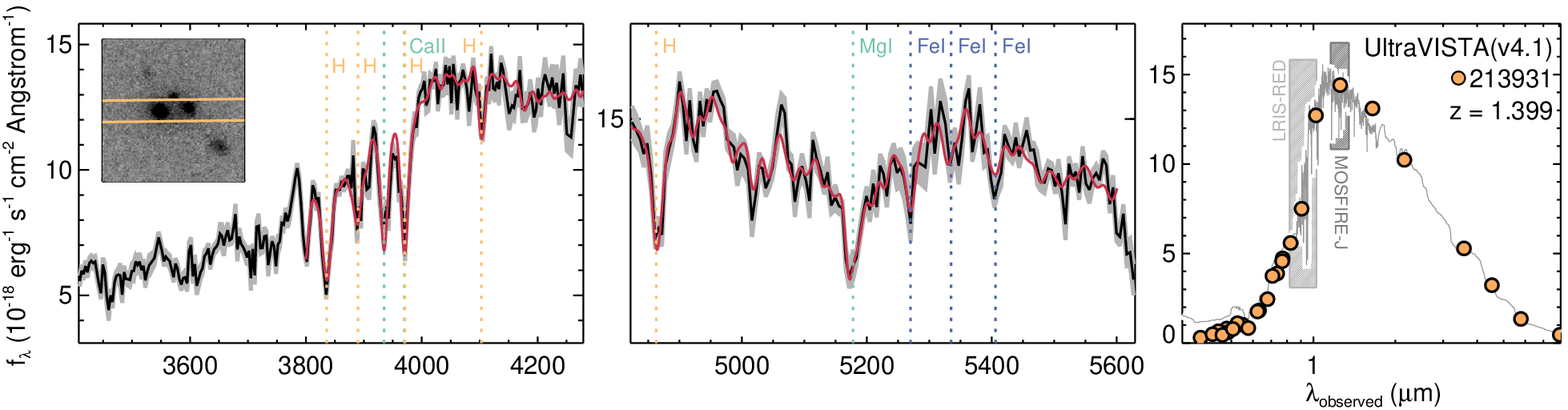}
  \includegraphics[width=0.95\textwidth]{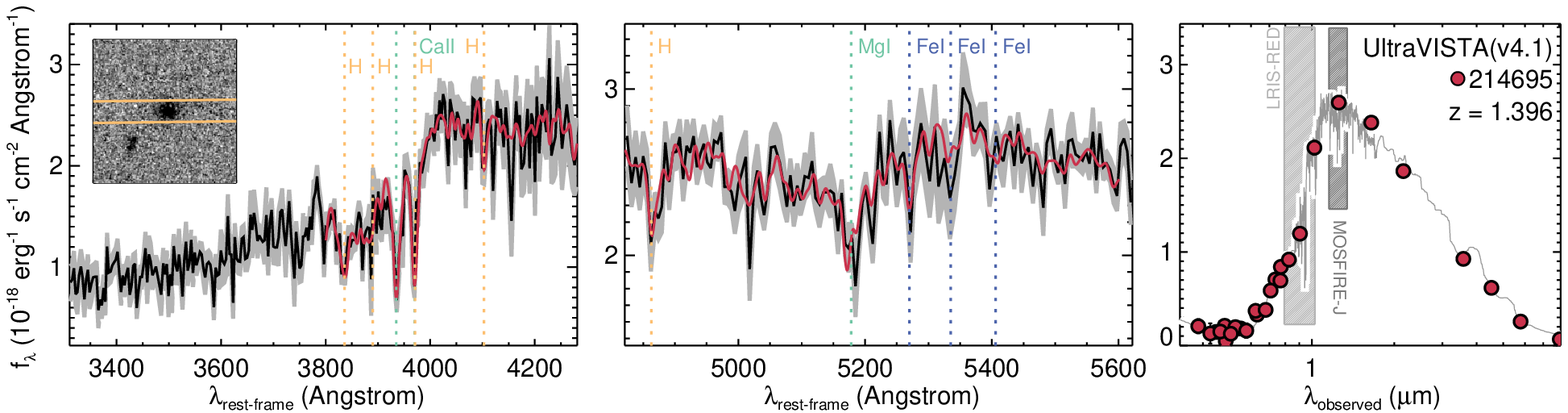}
\caption{{\it Left}: LRIS-RED and MOSFIRE/$J$-band spectra of 5 massive quiescent galaxies at $z\sim1.4$ (black). The spectra are binned by 10 pixels, corresponding to $\sim$3.3 and $\sim$5.4 rest-frame \AA\ per bin for the LRIS and MOSFIRE spectra, respectively. The gray areas represent the 1\,$\sigma$ uncertainty for the binned spectra. The best-fit \texttt{alf} models are shown in red. The F814W image (4\farcs5$\times$4\farcs5) is shown in the inset, with the MOSFIRE slit overplotted (91$^{\rm o}$, 0\farcs7 width). The LRIS slit had a similar orientation (93${\rm ^o}$) and width (1\arcsec). {\it Right:} UltraVISTA photometry (circles) and best-fit FSPS model (grey) for the same galaxies.\label{fig:spectra}}
\end{figure*}

The LRIS mask containing our five primary targets was observed for
$\sim$4.5\,hrs on 2017/01/04 using the 600/10000 red grating,
1\arcsec-wide slits, and an ABC dither pattern. The sky was clear with
an image quality of $\sim0\farcs8-1\farcs0$. The same five galaxies
were observed within one MOSFIRE mask on 2017/03/15 and 2017/04/5-6
for $\sim$12\,hrs in $J$, using 0\farcs7-wide slits and an ABA'B'
dither pattern \citep{MKriek2015}. The sky was clear and the seeing
varied between 0\farcs5-1\farcs0. For both instruments we assigned a
star to one of the slits to monitor weather conditions and aid the
data reduction.

The LRIS and MOSFIRE 2D spectra were reduced using custom
software. Initial sky subtraction was performed using the average of
the surrounding frames with the same integration times. Cosmic rays
were identified using L.A. Cosmic \citep{PvanDokkum2001} and combined
with a bad pixel map. Next, the individual sky-subtracted frames and
corresponding masks were resampled onto a common grid to account for
the wavelength calibration, dither position, distortions, and
drifts. An additional sky subtraction was performed by subtracting the
median background at each wavelength. For the MOSFIRE spectra, the
slit stars were used to derive a weighing factor for each science
exposure. All LRIS frames were weighted equally, since the weather
conditions were stable.

\begin{table}[t]
  \centering
  \caption{Parameters of $z\sim1.4$ Quiescent Galaxies}
  \label{tab:results}
  \begin{tabular}{l | c | r c c c c}
    \hline\hline
     \multicolumn{1}{c|}{ID\footnote{UltraVISTA catalog v4.1 by \cite{AMuzzin2013a}}} & \multicolumn{1}{c|}{\tt FAST} & \multicolumn{5}{c}{\tt alf}\\
     &  log\,$M$ & $\sigma_v$ & [Fe/H] & [Mg/Fe] & $t$ & $t_{\rm SSP}$\\
     &  $M_\odot$ & km/s & & & Gyr & Gyr \\
    \hline
   \bs  217249 & \bs 10.61 &  \bs $127^{+58}_{-29}$ &  \bs $-1.03^{+0.47}_{-0.27}$ & \bs $0.52^{+0.28}_{-0.35}$ & \bs $3.0^{+0.4}_{-0.6}$ & \bs $0.9^{+0.8}_{-0.4}$\\
    \bs 213947 &  \bs 10.87 &  \bs $170^{+25}_{-19}$ & \bs $-0.89^{+0.68}_{-0.24}$ & \bs $0.57^{+0.24}_{-0.32}$ & \bs $3.0^{+0.1}_{-1.0}$ & \bs $0.8^{+0.2}_{-0.3}$\\
   \bs  214340 &  \bs 10.80  &  \bs $79^{+24}_{-27}$ &  \bs $-0.42^{+0.16}_{-0.24}$ & \bs $0.22^{+0.19}_{-0.14}$ & \bs $3.8^{+5.3}_{-0.8}$ & \bs $1.4^{+0.6}_{-0.5}$\\
   \bs  213931\footnote{The mass for this galaxy represents the total mass, including all three clumps.} &  \bs 11.73  & \bs $342^{+12}_{-11}$ &  \bs $-0.27^{+0.07}_{-0.07}$ & \bs $0.44^{+0.08}_{-0.07}$ &  \bs $3.1^{+0.2}_{-0.1}$ &
     \bs $2.0^{+0.8}_{-0.5}$\\
   \bs  214695 & \bs 11.18  &  $209^{+30}_{-33}$ &  \bs $-0.20^{+0.17}_{-0.22}$ & \bs $0.28^{+0.15}_{-0.14}$ & \bs $4.5^{+2.9}_{-1.2}$ & \bs $3.8^{+7.3}_{-1.6}$ \\
    \hline \hline
  \end{tabular}
\end{table}

The individual spectra were average combined, taking into account both
the weighing factors and the corresponding rectified masks. The
relative flux calibration was performed using a response spectrum. For
MOSFIRE the response spectrum was derived from the spectra of A0\,V
stars. For LRIS we used the theoretical atmospheric absorption
spectrum, adjusted to match the atmospheric features in the spectrum of the slit star,
combined with the intrinsic shape of the slit star and other bright
objects in the mask. More details on the MOSFIRE reduction software,
which was developed for the MOSDEF survey, are given in
\cite{MKriek2015}. One-dimensional (1D) spectra were extracted using
an optimal weighing procedure. The combined LRIS and MOSFIRE spectra
for each galaxy, together with the HST F814W image
\citep{NScoville2007} and the photometric spectral energy distribution
\citep[][]{AMuzzin2013b} are shown in Figure~\ref{fig:spectra}. We
detect numerous stellar absorption lines for all five galaxies.

The most massive galaxy in the sample, 213931, consists of three
  separate clumps, which were blended together as one system in the
  UltraVISTA images and catalog. As illustrated in
Figure~\ref{fig:spectra}, the slit was aligned along the two most
massive clumps. With MOSFIRE we detect two blended traces, though one
of the traces is significantly fainter, and no separate analysis could
be performed. Using LRIS, we do not detect two separate traces, and
thus we treat 213931 as one system in our analysis. 

\section{Analysis}

We use the absorption line fitter ({\tt alf}) code to estimate
parameters from the combined 1D LRIS and MOSFIRE spectra. The code is
described in detail in \cite{CConroy2012d,CConroy2014,JChoi2014}. In
summery, {\tt alf} combines libraries of isochrones and empirical
stellar spectra with synthetic stellar spectra covering a wide range
of elemental abundance patterns. The code fits for C, N, O, Na, Mg,
Ca, Ti, V, Cr, Mn, Fe, Co, and Ni abundances, redshift, velocity
dispersion, stellar population age, and several emission lines.  The
ratio of the model and data is fit by a high order polynomial in order
to avoid potential issues with the flux calibration of the data. The
fitting is done using a Markov chain Monte Carlo (MCMC) algorithm
\citep{DForeman-Mackey2013}. This spectral modeling approach is
strongly preferred over the use of integrated absorption line
measurements for distant galaxies \citep[see discussion
  in][]{MKriek2016}. 

For our $z\sim1.4$ galaxies, we assume the \cite{PKroupa2001} IMF and
allow for a double-component stellar population with two different
ages. This model is preferred over a single-age model (i.e., SSP),
because it indirectly separates the younger and older stellar
populations and results in mass-weighted abundance measurements. In
case we assume an SSP, the younger stars will disproportionally
dominate the results.In Table~\ref{tab:results} we list
  mass-weighted abundance ratios and ages for the double-component model,
  as well as the ages for an SSP. Due to the mass weighing, the
  double-component ages are older than the SSP ages (which are closer
  to luminosity-weighted ages).  While all abundances are free
parameters, only few can be accurately determined. We also
  determine the total stellar metallicity using
  $\rm[Z/H]=[Fe/H]+0.94[Mg/Fe]$  \citep{DThomas2003}, and derive the
  16\% and 84\% confidence intervals from the [Z/H] distribution
  following from the MCMC simulations.

Stellar masses are derived by fitting the combined continuum spectra
and UltraVISTA photometry with the Flexible Stellar Population
Synthesis ({\tt FSPS}) models \citep{CConroy2009}, using the {\tt
  FAST} fitting code \citep{MKriek2009b}. We assume solar
  metallicity, a delayed exponential star-formation history, the
\cite{GChabrier2003} IMF, and the average attenuation law by
\cite{MKriek2013}. We caution though that these stellar masses
  are likely underestimated, and more complicated star formation
  histories, like the one assumed in the \texttt{alf} fitting, would
  lead to masses that are $\sim50\%$ larger.

Rest-frame optical sizes are derived from the combination of F814W
data from \cite{NScoville2007} and F160W data from COSMOS-DASH,
following the procedure described in \cite{LMowla2018}. This procedure
relies on S\'ersic fits and the \texttt{GALFIT} modeling code
\citep{CPeng2002}. We derive the (non-circularized) size at
5000\,\AA\ using a linear interpolation between the F814W and F160W
sizes. For 213931, we measure the size of the brightest clump.

\begin{figure*}[!tph]
\centering
\includegraphics[width=0.4\textwidth]{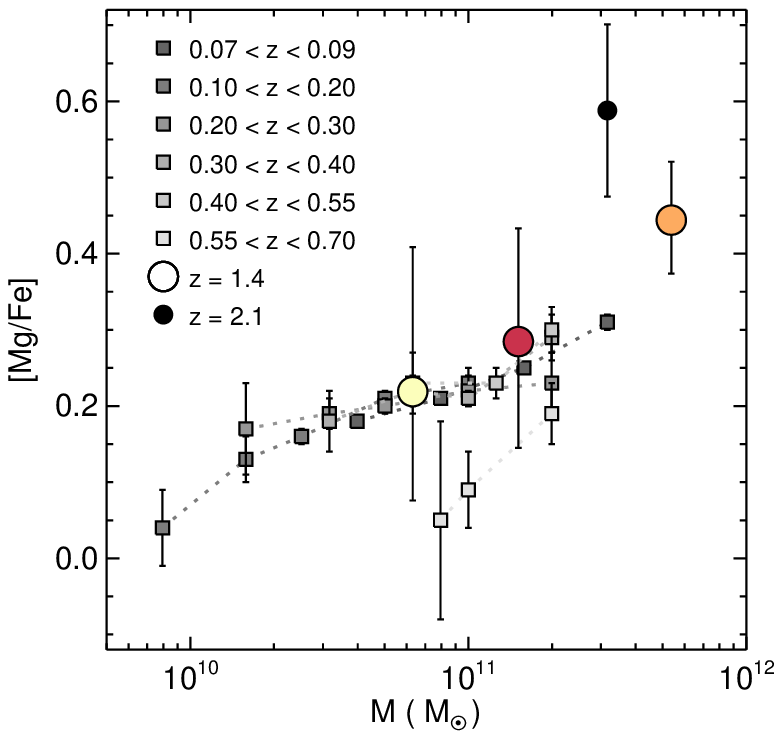}
\includegraphics[width=0.4\textwidth]{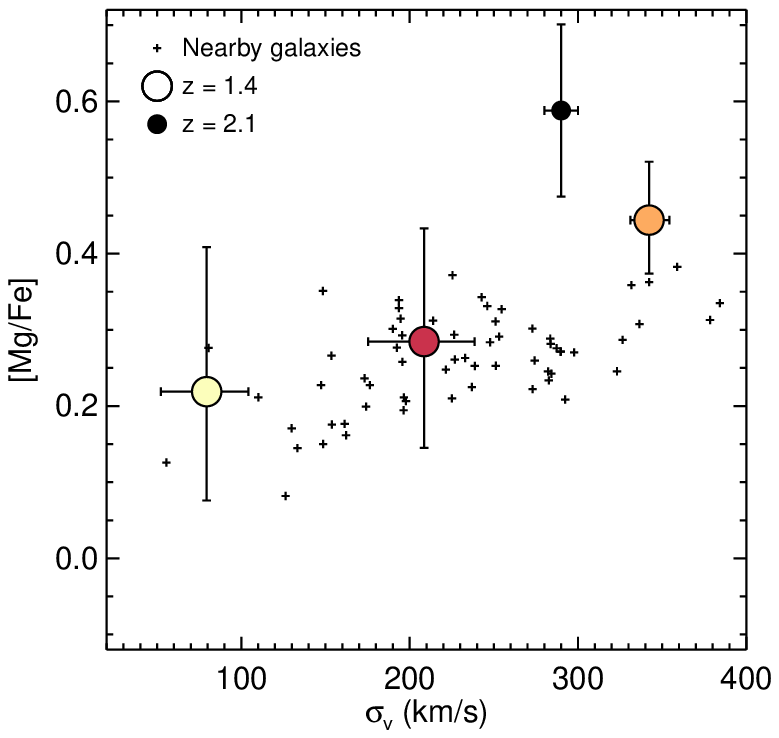}
\includegraphics[width=0.4\textwidth]{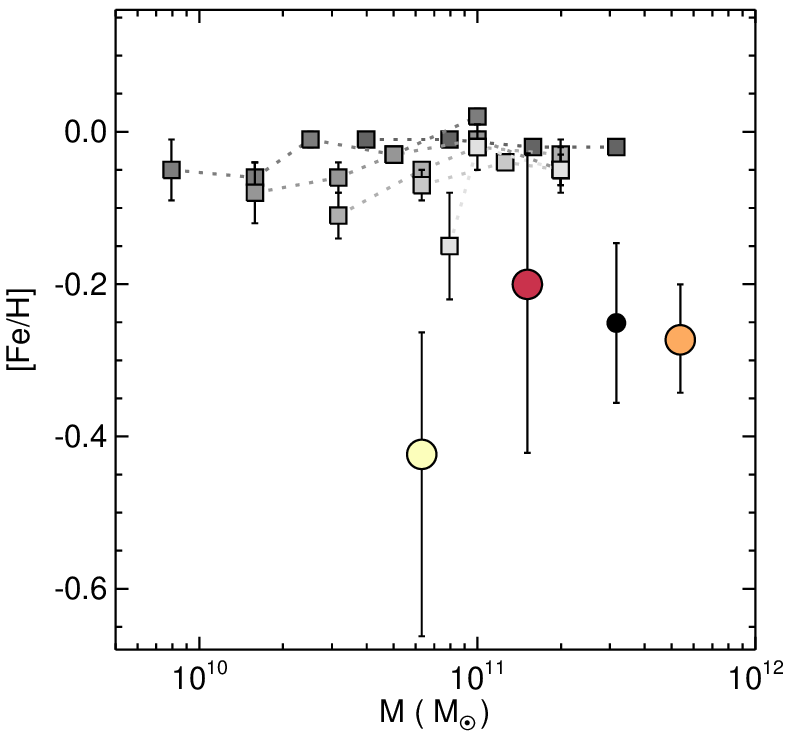}
\includegraphics[width=0.4\textwidth]{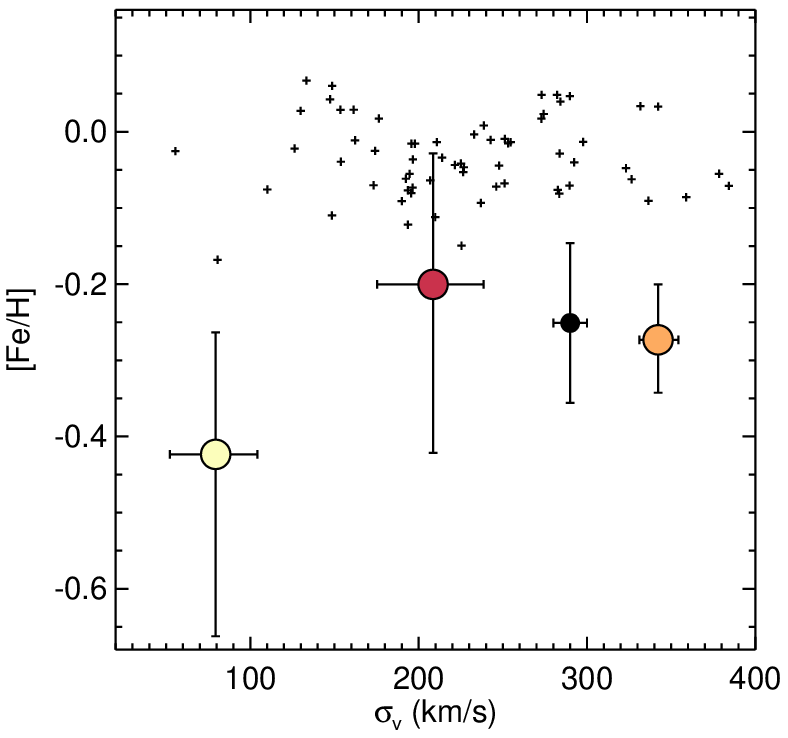}
\includegraphics[width=0.4\textwidth]{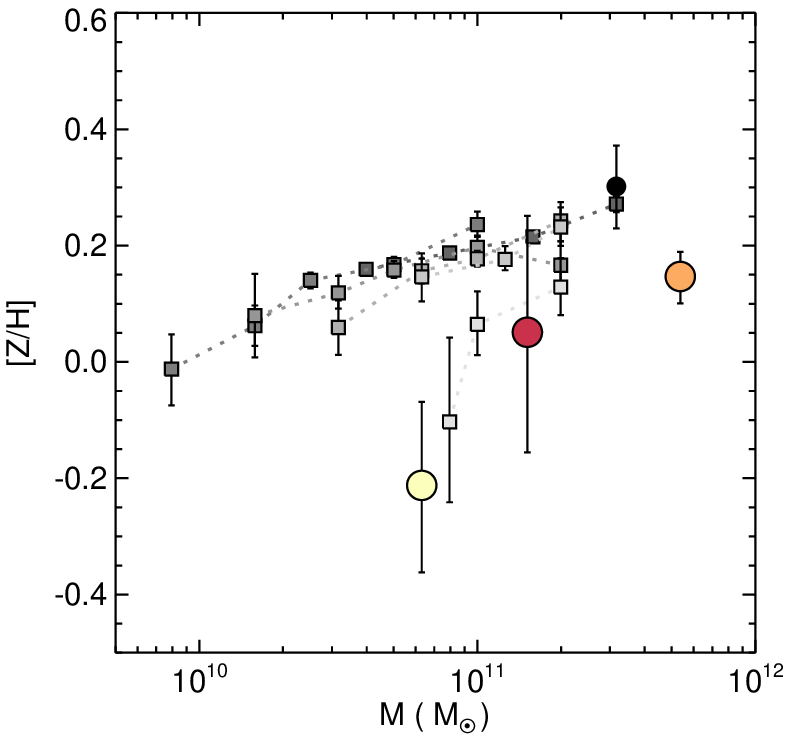}
\includegraphics[width=0.4\textwidth]{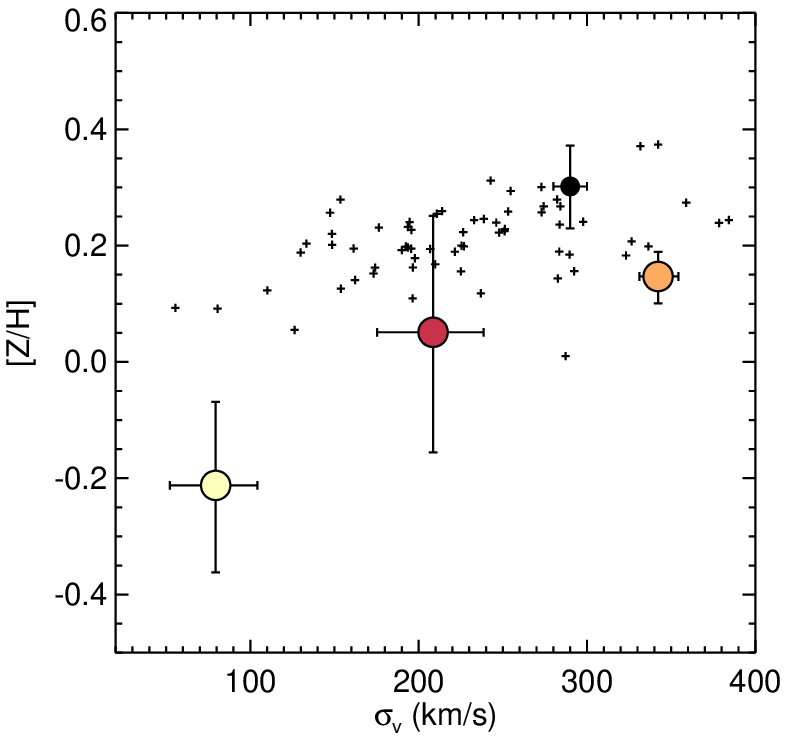}

\caption{[Mg/Fe] (top), [Fe/H] (middle) and [Z/H] (bottom) vs stellar
  mass (left) and velocity dispersion (right) of the $z\sim1.4$
  quiescent galaxies 214340 (yellow), 213931 (orange), and 214695
  (red), in comparison to nearby early-type galaxies
  \citep[plusses;][]{CConroy2012d}, stacks of quiescent galaxies at
  $0.07<z<0.70$ \citep[squares;][]{JChoi2014}, and a quiescent galaxy
  at $z=2.1$ \citep[black filled circle;][]{MKriek2016}. For all
  (stacks of) galaxies, the abundance ratios and velocity dispersions
  were derived using \texttt{alf}. The relations at $z\sim1.4$ are
  similar to those at $z<0.7$, though [Fe/H] seems offset to lower
  values. The bottom panels show that a tentative stellar
  mass-metallicity relation may already be in place at $z\sim1.4$,
  possibly offset to lower metallicities.  \label{fig:mgfeh_ms}}

\end{figure*}

\section{Results}

In Figure~\ref{fig:mgfeh_ms}, we show the abundance ratios [Mg/Fe] and
[Fe/H] of the $z\sim1.4$ galaxies in comparison with galaxies at lower
redshift \citep{JChoi2014,CConroy2012d} and a single massive quiescent
galaxy at $z=2.1$ \citep{MKriek2016}. Although metal and Balmer
absorption lines are detected for all five galaxies, the abundance
ratios for 217249 and 213947 are poorly
constrained. Figures~\ref{fig:uvj} and \ref{fig:spectra} show that
these galaxies have blue rest-frame colors, strong Balmer lines,
and are dominated by A-type stars. Hence, their luminosity-weighted
ages are relatively young, and thus their metal lines are 
weak. Therefore, in Figure~\ref{fig:mgfeh_ms} we only show the three
redder quiescent galaxies at $z\sim1.4$. For these galaxies we
find similar abundance ratios when adopting a single-age model. 

In the left panels of Figure~\ref{fig:mgfeh_ms}, we show the abundance
ratios [Mg/Fe] and [Fe/H] versus stellar mass. We compare our results
to the work by \cite{JChoi2014}, based on stacked spectra of quiescent
galaxies at $0.07<z<0.7$. In the right panels we show the abundance
ratios as a function of the observed integrated stellar velocity
dispersion ($\sigma_v$), in comparison to the nearby early-type galaxy
sample by \cite{CConroy2012d}. All abundance ratios and velocity
dispersions have been derived using the \texttt{alf} code. However,
stellar masses and velocity dispersions were not consistently derived
for all samples, and thus we only show the comparison samples if the
measurements are available.

Similar to $z<0.7$, [Mg/Fe] appears positively correlated with both
stellar mass and $\sigma_v$ at $z\sim1.4$. These results, however, are
based on only three galaxies; a larger sample is needed to confirm
these trends. The two lower-mass galaxies (214340 \& 214695) have
[Mg/Fe] similar to their lower-redshift analogs. With a [Mg/Fe] of
0.44, the most massive galaxy (213931) is slightly offset from the
low-redshift [Mg/Fe]-mass and [Mg/Fe]-$\sigma_v$ relations. This
offset, however, is not significant, and larger galaxy samples are
needed to assess whether the most massive galaxies are indeed more
$\alpha$-enhanced at earlier times.  Similar to the low-redshift
samples, we find no correlation between [Fe/H] and mass or
$\sigma_V$. We do find an offset in [Fe/H], such that the $z\sim1.4$
galaxies are deficient in iron compared to similar-mass low-redshift
galaxies.

\begin{figure}
\centering
\includegraphics[width=0.4\textwidth]{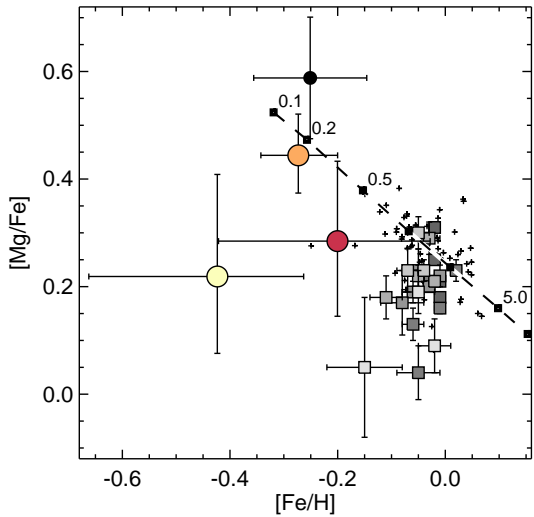}
\caption{[Mg/Fe] vs [Fe/H] for massive quiescent galaxies at
  $z\sim1.4$ (colored symbols), in comparison to nearby early-type
  galaxies (plusses), stacks of quiescent galaxies at $0.07<z<0.70$
  (gray squares), and a quiescent galaxy at $z=2.1$ (black filled
  circle). Symbols are the same as in Figure~\ref{fig:mgfeh_ms}. The
  dashed line represents a chemical evolution model, with different
  star-formation timescales in Gyr indicated by the small black
  squares. The abundance ratios of the two most massive galaxies
  (orange and black circles) at $z\gtrsim1.4$ suggest that
  their star-formation timescales are shorter than for $z<0.7$
  galaxies.}
\label{fig:mgfe_feh}
\end{figure}

In Figure~\ref{fig:mgfe_feh} we compare the combined abundance ratios
[Mg/Fe] and [Fe/H] to a closed-box chemical evolution model for
different star-formation timescales. This model assumes a
\cite{ESalpeter1955} IMF, a constant SFR, the core-collapse and Type
Ia supernova yield models by \cite{CKobayashi2006} and
\cite{KNomoto1984}, and a Type Ia delay time distribution of the form
$t^{-1}$ between 0.1 and 13~Gyr \citep{DMaoz2012}. Comparison with
this model indicates that the $z\sim1.4$ galaxies formed their stars
over a brief time period of $\sim0.2-1$~Gyr. As discussed in
\cite{MKriek2016}, by adopting different core-collapse supernova
yields or a different Type Ia delay time distribution, we can change
this model substantially.

\section{Discussion}\label{sec:dis}

In this Letter, we present ultradeep rest-frame optical spectra of
five massive quiescent galaxies at $z\sim1.4$, all of which show
multiple stellar absorption lines. For three galaxies we derive the
abundance ratios [Mg/Fe] and [Fe/H], but the remaining two galaxies
have too young luminosity-weighted ages to yield robust
measurements. Similar to $z<0.7$ studies, we find a tentative positive
relation between [Mg/Fe] and stellar mass (or velocity
dispersion). Also similar to $z<0.7$, we find no correlation between
[Fe/H] and stellar mass (or velocity dispersion). Our results
  may imply that the stellar mass-metallicity relation was already in
  place at $z\sim1.4$ (bottom panels, Fig.~\ref{fig:mgfeh_ms}).

While the [Mg/Fe]-mass relation at $z\sim1.4$ is consistent with the
$z\lesssim0.7$ relation, [Fe/H] at $z\sim1.4$ is $\sim$0.2~dex lower
than at $z<0.7$. We also found a low [Fe/H] for a single massive
quiescent galaxy at $z=2.1$ \citep{MKriek2016}. In addition to the low
[Fe/H], this $z=2.1$ galaxy was more $\alpha$-enhanced than
low-redshift galaxies of similar mass, with a [Mg/Fe] of 0.6. The most
massive galaxy in the current sample has a [Mg/Fe] of
$0.44^{+0.08}_{-0.07}$, and thus may also be more $\alpha$-enhanced
than lower-redshift analogs. Combined, these results may suggest that
[Mg/Fe] of the most massive galaxies decreases over cosmic time,
possibly by accretion of low-mass, less $\alpha$-enhanced galaxies.

A similar scenario has been proposed to explain the evolution in the
mass-size relation of quiescent galaxies between $z\sim2$ and $z\sim0$
\citep[e.g.,][]{PvanDokkum2008,RBezanson2009,TNaab2009}. In this
context, we note that all three $z\sim1.4$ galaxies as well as the
$z\sim2.1$ galaxy have close neighbors, and thus may be in the process
of merging with smaller galaxies \citep[see also][]{MGu2018}.
However, as [Fe/H] is constant with mass, it is not obvious how this
scenario could increase [Fe/H], and a larger galaxy sample is needed
to understand the evolutionary scenario \citep{JChoi2014}. Other
possible scenarios include late-time star formation and the growth of
the quiescent galaxy population over time. Galaxies that stop forming
stars at later times will have lower [Mg/Fe], higher [Fe/H], and
larger sizes \citep[][]{SKhochfar2006}. Thus, once these galaxies join
the quiescent population, they will alter the average size and
abundance ratios \citep[e.g.,][]{MCarollo2013,JChoi2014}. 

Our metallicities are higher than the sub-solar metallicities found by
\cite{TMorishita2018} for two $10^{11}~M_\odot$ quiescent galaxies at
$z\sim2.2$, but lower than the supersolar metallicities found for a
stack of massive (log\,$M/M_\odot\sim11.4$) galaxies at $z\sim1.6$
\citep[{[Z/H]}$=0.24^{+0.20}_{-0.14}$;][]{MOnodera2015}, our $z=2.1$ galaxy
\citep[{[Z/H]}$=0.30\pm0.07$;][]{MKriek2016}, as well as the $z\sim1.4$ galaxy
by \citet[][{[Z/H]}$>$0.5]{ILonoce2015}. Our results are more similar
to the solar metallicities found by \cite{VEstrada2018} for stacks of
$1.0<z<1.8$ galaxies (log$M/M_\odot\sim10.8$). We do caution though
that these studies use varying datasets and techniques to measure
stellar metallicities, and thus different assumptions and other
systematics complicate the comparison. 

Larger samples are needed to obtain a full census of the relation
between the chemical composition, stellar mass, and structures of
distant quiescent galaxies. This work is part of a survey to
obtain deep rest-frame optical spectra and elemental abundance
measurements for a sample of 20 distant quiescent galaxies; 10
galaxies at $z\sim1.4$ and 10 galaxies at $z\sim2.1$. The current
study demonstrates that such measurements are only possible for
individual quiescent galaxies that are dominated by an older stellar
population, which have more pronounced metal lines. For quiescent
galaxies that are dominated by a younger stellar population we have to
rely on stacking techniques to obtain more robust measurements. In the
near future, we will use our full sample to measure the relations
between the chemical composition, age, stellar mass/velocity
dispersion, and galaxy sizes/structures at $z\sim1.4$ and $z\sim2.1$,
and compare them to the lower redshift relations. In the more distant
future, NIRSpec on JWST will enable the first resolved stellar
abundance studies at these redshifts. These two studies together, in
combination with the resolved stellar abundance studies at low
redshift \citep[e.g.,][]{JGreene2015}, will open a new
window into the chemical enrichment, star formation, and assembly
histories of massive quiescent galaxies. 

\acknowledgements 

We thank the referee for a very thoughtful and constructive report,
Ryan Trainor for his help with the LRIS observations, and Jenny Greene
and Marijn Franx for valuable discussions. We acknowledge support from
NSF AAG grants AST-1908748 and 1909942. The authors wish to recognize
and acknowledge the very significant cultural role and reverence that
the summit of Mauna Kea has always had within the indigenous Hawaiian
community. We are most fortunate to have the opportunity to conduct
observations from this mountain.


\end{document}